\documentclass[english]{aastex6}
\usepackage[T1]{fontenc}
\usepackage[latin9]{inputenc}
\usepackage{array}
\usepackage{bm}
\usepackage{multirow}
\usepackage{graphicx}
\usepackage{esint}

\makeatletter

\providecommand{\tabularnewline}{\\}

\usepackage{aas_macros}
\usepackage{subfigure}

\makeatother

\usepackage{babel}
\begin{document}
\global\long\def\grad{\bm{\nabla}}

\title{Strategies in seismic inference of supergranular flows on the Sun}

\author{Jishnu Bhattacharya \& Shravan M. Hanasoge}

\affil{Department of Astronomy and Astrophysics, Tata Institute of Fundamental
Research, Mumbai-400005, India}
\begin{abstract}
Observations of the solar surface reveal the presence of flows with
length scales of around $35$ Mm, commonly referred to as supergranules.
Inferring the sub-surface flow profile of supergranules from measurements
of the surface and photospheric wavefield is an important challenge
faced by helioseismology. Traditionally, the inverse problem has been
approached by studying the linear response of seismic waves in a horizontally
translationally invariant background to the presence of the supergranule;
following an iterative approach that does not depend on horizontal
translational invariance might perform better, since the misfit can
be analyzed post iterations. In this work, we construct synthetic
observations using a reference supergranule, and invert for the flow
profile using surface measurements of travel-times of waves belonging
to modal ridges $f$ (surface-gravity) and $p_{1}$ through $p_{7}$
(acoustic). We study the extent to which individual modes and their
combinations contribute to infer the flow. We show that this method
of non-linear iterative inversion tends to underestimate the flow
velocities as well as inferring a shallower flow profile, with significant
deviations from the reference supergranule near the surface. We carry
out a similar analysis for a sound-speed perturbation and find that
analogous near-surface deviations persist, although the iterations
converge faster and more accurately. We conclude that a better approach
to inversion would be to expand the supergranule profile in an appropriate
basis, thereby reducing the number of parameters being inverted for
and appropriately regularizing them.
\end{abstract}

\section{Introduction}

Helioseismology has allowed us to probe sub-surface features that
are otherwise inaccessible to electromagnetic observations of the
solar surface. Seismology attempts to uncover flows beneath the photosphere;
such estimates act as complementary constraints on simulations of
convective flows and solar dynamics. A moving background medium would
advect seismic waves, and the primary impact is expected to be a shift
in measured travel-time. Tracking a propagating seismic wave \citep{duv93,duv97}
would therefore allow us to draw inferences about local sub-surface
flows such as supergranules. Several authors \citep{kos00,gb02} have
formulated inverse problems to relate measured travel-times of seismic
waves on the solar surface to the shape and structure of the flow
profile underneath. Attempts at reconstructing synthetic flows using
seismic measurements, e.g. by \citet{dom13}, however, suggest that
it is difficult to capture vertical flows accurately, which they have
attributed to a combination of regularization parameter, signal cross-talk,
inconsistent boundary conditions. While these issues are certainly
important, they must be disentangled from limitations arising due
to a limited set of spectral modes and linear inversions.

A perturbation such as supergranular flow leaves its imprint on propagating
seismic waves; the difference in surface measurements between a model
that accounts for the perturbation and one that doesn't can be represented
in terms of a misfit functional $\chi$. The dominant effect of flow
velocities on seismic waves is expected to be a shift in measured
arrival times of wavepackets, and therefore one possible misfit functional
$\chi$ that captures the impact of flows on seismic waves is the
$L_{2}$ norm of the difference in time of arrival of a wave at different
points on the surface, in various models of the supergranular flow
profile. We denote spatial locations in the Sun by the coordinate
$\mathbf{x}$, where the bold-face notation refers to vectors. Assuming
that the observed arrival time $\tau_{i}^{o}$ at the point $\mathbf{x}_{i}$
differs from the time $\tau_{i}$ predicted by a model of the flow,
the misfit $\chi$ can be expressed as 
\begin{equation}
\chi=\frac{1}{2}\sum_{\mathbf{x}_{i}}\left(\tau_{i}-\tau_{i}^{o}\right)^{2},
\end{equation}
where the travel time $\tau_{i}$ predicted by the model depends on
specifics of the model's internal structure, in this case the flow
velocity as a function of space. We ignore noise in this analysis,
but the definition can be easily extended to account for it. We represent
these structural parameters of the model as $\mathbf{m}\left(\mathbf{x}\right)$.
If the model changes by an amount $\delta\mathbf{m}\left(\mathbf{x}\right)$,
the misfit changes by 
\begin{equation}
\delta\chi=\int d\mathbf{x}\,\left[\sum_{\mathbf{x}_{i}}\left(\tau_{i}-\tau_{i}^{o}\right)\frac{\partial\tau_{i}}{\partial\mathbf{m}\left(\mathbf{x}\right)}\right]\delta\mathbf{m}\left(\mathbf{x}\right)+\mathcal{O}\left(\left\Vert \delta\mathbf{m}^{2}\right\Vert \right).\label{eq:misfit_linear}
\end{equation}
The extent to which surface measurements of seismic wave travel-times
are sensitive to an update in the model at some spatial location $\mathbf{x}$
is referred to as the source-receiver ``sensitivity kernel'', and
is given by 
\begin{equation}
K_{SR}\left(\mathbf{x};\mathbf{x}_{i},\mathbf{m}\left(\mathbf{x}\right)\right)=\frac{\partial\tau_{i}}{\partial\mathbf{m}\left(\mathbf{x}\right)}.
\end{equation}
The kernel depends on the location of the wave source that is implicit
in the present analysis, the location of the measurement point $\mathbf{x}_{i}$
and also on the background model $\mathbf{m}\left(\mathbf{x}\right)$
about which we compute the linear expansion. Each measurement point
$\mathbf{x}_{i}$ contributes its own kernel, and the weighted sum
that governs the change in misfit is referred to as an ``event'' kernel
--- 
\[
K\left(\mathbf{x};\mathbf{m}\left(\mathbf{x}\right)\right)=\sum_{\mathbf{x}_{i}}\left(\tau_{i}-\tau_{i}^{o}\right)\frac{\partial\tau_{i}}{\partial\mathbf{m}\left(\mathbf{x}\right)}.
\]

Seismic inverse problems have been traditionally posed by computing
the kernel $K$ about a horizontally translationally invariant model
$\mathbf{m}\left(\mathbf{x}\right)$. \citep[eg.][]{gb02,dom13,sva11}.
Such a formulation has the advantage of allowing spectral decomposition
horizontally to substantially simplify the inversion process; as a
corollary, it treats the change in the model $\delta\mathbf{m}$ as
a one-time update that minimizes the misfit $\chi$. The approach
is fruitful if the perturbation $\delta\mathbf{m}$ is ``small''
in some sense that justifies the truncation to linear order in Equation
(\ref{eq:misfit_linear}); however it is not clear if such one-step
inversions will always suffice. \citet{jack07} showed that in presence
of uniform flows with velocity above $200$ m/s, the deviations in
temporal cross-covariance of $f$-mode waveforms could only be reproduced
sufficiently accurately by including terms up to third order in velocity,
an approach that goes far beyond the linear approximation presented
above. Supergranular velocities are expected to breach $300$ m/s
\citep{hath00}, therefore lying in the \citet{jack07} regime; however
sensitivity kernels incorporating third order corrections are expensive
to compute in general, even in spectral space. One way to circumvent
this problem is by carrying out an iterative inversion, as was demonstrated
by \citet{han14} (referred to as H14 hereafter), where one recomputes
the sensitivity kernel $K\left(\mathbf{x};\mathbf{x}_{i},\mathbf{m}\left(\mathbf{x}\right)\right)$
after each small step $\delta\mathbf{m}$ in model space. This approach
relaxes the assumption of horizontal translation invariance in the
model $\mathbf{m}\left(\mathbf{x}\right)$, with the caveat that one
must recourse to evaluating the kernel in position space instead of
a spectral basis. The advantage of this approach is that one can iteratively
update the model, and therefore each individual change in the model
may be small and lie within the linear regime required for Equation
(\ref{eq:misfit_linear}) to hold. Another advantage is that one can
examine the model at each stage during the iterations to analyze the
impact of inversion strategies such as spectral filtering and regularization
on the final inference.

In this paper, we carry out an analysis similar to H14; we choose
a reference supergranule and iteratively improve a flow profile based
on the misfit between simulated surface wavefields in the two models.
H14 had pointed out that global seismology does a much better job
than local seismology at radially localizing small perturbations,
because of the larger number of radial orders identified. A question
arises naturally --- to what extent would the local seismic inferences
improve if more radial orders are included in the inversion? The original
analysis by H14 was restricted to $f$ and $p_{1}$ modes; in this
work we increase the domain of measurements used in the inversion
process by (a) including radial orders up to $p_{7}$ (b) computing
wave travel-times over a larger range of distances for each ridge
(c) studying the impact of different levels of regularization on the
kernel. We systematically study the misfit in each radial order, and
investigate the extent to which they contribute towards the iterated
flow profile. This analysis allows us to appreciate the shortcomings
of such an approach in inferring sub-surface flows, and drives us
along the trajectory towards better strategies leading to accurate
inversions.

\section{Problem Setup}

Supergranules on the Sun exhibit spatial scales ranging from $20$
Mm to $75$ Mm \citep{rieu08}; the power spectrum of Doppler velocity
peaks at around $35$ Mm \citep{hath00,rieu08,duv10}. While velocity
fields on the surface can be measured accurately, the sub-surface
profile of the flow is uncertain. Inferred models of sub-surface flow
profile differ based on the inversion technique used. \citet{duvhan13}
(hereafter DH2) had studied $f$-mode travel-time shifts produced
by an ``averaged supergranule'', and had suggested the use of an
axisymmetric model with peak radial velocity $697\,\mathrm{m/s}$
at a depth of $-1.6$ Mm below the surface, and a peak vertical velocity
of $240\,\mathrm{m/s}$ at a depth of $-2.3$ Mm. In a recent work,
however, \citet{degrave15} showed that while the DH2 model produced
ray-theoretic travel-times similar to measured ones, there was a notable
discrepancy with travel-times computed using Born kernels. Within
the qualitative profile of the DH2 model, the authors proposed two
best-fit models: one having lower velocity magnitudes and peaking
at a shallower depth; the other having larger velocities and extending
deeper. Our aim in this work is to validate our inversion algorithm
for one plausible model of supergranular flow, hence we choose to
work with DH2, but it is straightforward to extend our analysis to
other models. 

We work in 2D Cartesian coordinates denoted by $\mathbf{x}=\left(x,z\right)$,
with the vertical $z$ axis representing the direction along the line
of sight, and the horizontal $x$ axis lying along a direction perpendicular
to $z$. We assume that the background solar model is stratified vertically,
and may be parameterized in terms of the density $\rho\left(z\right)$,
sound-speed $c\left(z\right)$, pressure $p\left(z\right)$, acceleration
due to gravity $\mathbf{g}\left(z\right)=-g\left(z\right)\,\hat{\mathbf{e}}_{z}$.
In the subsequent analysis we suppress the explicit coordinate dependence
to simplify notation wherever it is unambiguous. We neglect variations
in the background thermal structure associated with flows, noting
that their impact on travel-times is not expected to be significant
\citep{han12}. We choose a convectively stabilized variant of Model
S \citep{jcd1996,handuvcou2007} as our background. On top of this,
we implant a two-component supergranular flow profile $\mathbf{v}\left(\mathbf{x}\right)=\left(v_{x}\left(\mathbf{x}\right),v_{z}\left(\mathbf{x}\right)\right)$
that is stationary in time. We enforce mass conservation
\begin{equation}
\grad\cdot\left(\rho\mathbf{v}\right)=0,
\end{equation}
where $\grad$ is the spatial derivative, and start from a vector
potential $\psi$ that is related to the flow through
\begin{equation}
\mathbf{v}^{\mathrm{ref}}=\frac{1}{\rho}\grad\times\left[\rho c\psi^{\mathrm{ref}}\mathbf{e}_{y}\right],\label{eq:v_psi}
\end{equation}
In this formulation the vector potential $\psi^{\mathrm{ref}}$ has
the dimension of length.\textbf{ }A specific functional form of $\psi^{\mathrm{ref}}$
had been proposed by \citet{duvhan13} to match the $10\,\mathrm{m/s}$
vertical flow at the photosphere as measured by \citet{duv10}, and
is given by

\textbf{
\begin{eqnarray}
\psi_{DH2}^{\mathrm{ref}} & = & \frac{v_{0}}{c\left(z\right)}\frac{1}{k}J_{1}\left(k\left|x\right|\right)\times\nonumber \\
 &  & \exp\left(-\frac{\left(z-z_{0}\right)^{2}}{2\sigma_{z}^{2}}-\frac{\left|x\right|}{R}\right),\label{eq:dh2}
\end{eqnarray}
}where $J_{1}\left(x\right)$ is the Bessel function of order $1$
and argument $x$, and constants $v_{0}$, $z_{0}$ and $\sigma_{z}$
are fixed by comparing with measured surface velocities and wave travel-times.
It must be noted that the original analysis by \citet{duvhan13} was
in polar coordinates, hence the vector $\psi^{\mathrm{ref}}$ would
point along the azimuthal direction $\hat{\phi}$; we have represented
the functional form of the potential in Cartesian coordinates. For
the purpose of our analysis, however, we require mass conservation
in Cartesian coordinates, and therefore a potential directed along
$\hat{y}$ that would produce velocity profiles qualitatively similar
to those obtained from Equation (\ref{eq:dh2}). To this end, we modify
the profile slightly to obtain 
\begin{eqnarray}
\psi^{\mathrm{ref}} & = & \frac{v_{0}}{c\left(z\right)}\frac{\mathrm{sign}\left(x\right)}{k}\,J_{1}\left(k\left|x\right|\right)\times\nonumber \\
 &  & \exp\left(-\frac{\left(z-z_{0}\right)^{2}}{2\sigma_{z}^{2}}-\frac{\left|x\right|}{R}\right),
\end{eqnarray}
where the constant parameters are $v_{0}=240\,\mathrm{ms^{-1}}$,
$k=2\pi/30\,\mathrm{rad\,Mm^{-1}}$, $R=7.5\,\mathrm{Mm}$, $z_{0}=-2.3\,\mathrm{Mm}$
and $\sigma_{z}=0.912\,\mathrm{Mm}$. We use this as our reference
supergranule model to construct synthetic observations that we use
in our inversion. The supergranular flow velocity profile corresponding
to this vector potential is given by 
\begin{eqnarray}
v_{x} & = & v_{0}\frac{\mathrm{sign}\left(x\right)}{k}\,J_{1}\left(k\left|x\right|\right)\left(\frac{\left(z-z_{0}\right)}{\sigma_{z}^{2}}-\frac{\rho^{\prime}\left(z\right)}{\rho\left(z\right)}\right)\nonumber \\
 &  & \times\exp\left(-\frac{\left(z-z_{0}\right)^{2}}{2\sigma_{z}^{2}}-\frac{\left|x\right|}{R}\right),\\
v_{z} & = & v_{0}\left(\frac{1}{2}\left(J_{0}\left(k\left|x\right|\right)-J_{2}\left(k\left|x\right|\right)\right)-\frac{1}{kR}J_{1}\left(k\left|x\right|\right)\right)\nonumber \\
 &  & \times\exp\left(-\frac{\left(z-z_{0}\right)^{2}}{2\sigma_{z}^{2}}-\frac{\left|x\right|}{R}\right).
\end{eqnarray}
One notable departure from the DH2 model is that in our case, the
vertical component of velocity is lower than DH2 by a factor of $2$,
having a peak magnitude of $120\,\mathrm{m/s}$. While this is not
ideal, it is not an issue for us since the inversion algorithm can
be validated against any plausible model. The horizontal component
of velocity is similar to that obtained from DH2.
\begin{figure}
\begin{centering}
\includegraphics[scale=0.5]{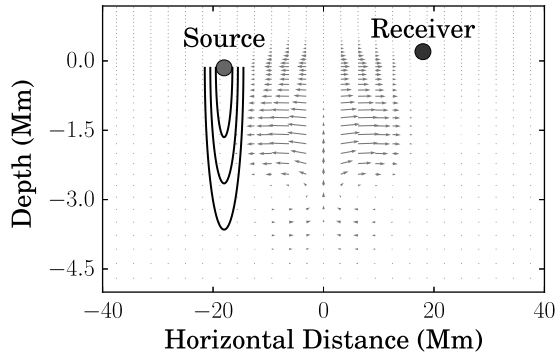}
\par\end{centering}

\caption{Schematic depiction of the measurement process used in our analysis.
Seismic waves are emitted by the source and travel through the supergranule.
The wave displacement and velocity are measured at various points
on the surface referred to as ``receivers''. We use eight different
sources spread out horizontally to adequately sample the supergranule
flow profile. \label{fig:flow_src}}
\end{figure}
\begin{figure}
\begin{centering}
\includegraphics[scale=0.5]{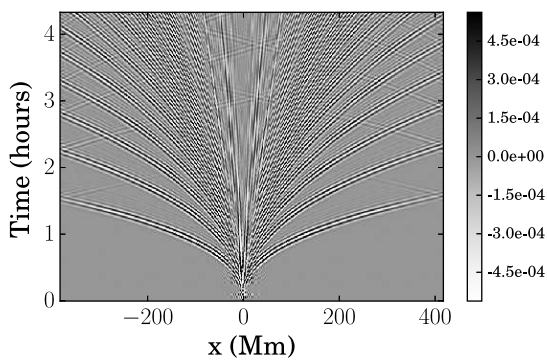}
\par\end{centering}

\caption{Vertical velocity of a wave emitted by a source at $x=0$ Mm as measured
at the surface. Receivers at each horizontal location record the arrivals
of seismic waves. Wavepackets reach receivers in discrete bunches,
corresponding to total internal reflection at different depths. \label{fig:Time-distance-diagram}}
\end{figure}
\begin{figure}
\begin{centering}
\includegraphics[scale=0.5]{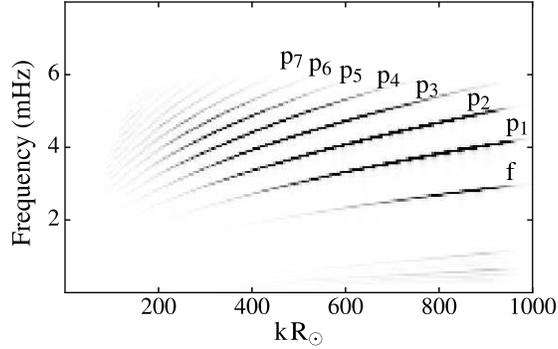}
\par\end{centering}

\caption{Spectrum of wave displacement as measured on the surface. The ridges
represent radial orders that probe different depths inside the Sun.
We use ridges \emph{f} through $p_{7}$ in our analysis.\label{fig:spectrum}}
\end{figure}

Seismic waves traveling through the background medium are advected
by the flow field $\mathbf{v}\left(\mathbf{x}\right)$ in addition
to experiencing restoring forces by pressure perturbations and gravity.
The linearized wave equation for the small-amplitude displacement
$\bm{\xi}\left(\mathbf{x},t\right)$ is given by 
\begin{eqnarray}
\rho\partial_{t}^{2}\bm{\xi}+2\rho\mathbf{v}\cdot\grad\partial_{t}\bm{\xi} & = & \grad\left(c^{2}\rho\grad\cdot\bm{\xi}+\bm{\xi}\cdot\grad p\right)\nonumber \\
 &  & +\mathbf{g}\grad\cdot\left(\rho\bm{\xi}\right)+\mathbf{S},
\end{eqnarray}
where $\mathbf{S}=\mathbf{S}\left(\mathbf{x},t\right)$ represents
sources that are exciting waves. We choose eight sources that are
isolated and fire independently of each other. The fundamental measurement
that we consider is the vertical velocity field of waves measured
at the surface. Each spatial point at which a measurement is made
is termed a ``receiver''. Observations of seismic waves on the solar
surface would measure source-receiver travel-times in the presence
of supergranules. An incomplete model of the supergranule flow profile
would produce travel-times that differ from the observed ones. The
idea of full-waveform inversion, as prescribed by \citet{hantromp14}
in this context, is to iteratively reduce the misfit between the observed
and predicted travel-times by altering the flow velocity in the model
at each step. Denoting the travel-time measured at receiver location
$\mathbf{x}_{i}$ in the model with the reference supergranule by
$\tau_{i}^{o}$, and that measured in the iterated model by $\tau_{i}$,
we define a least-square misfit functional 
\begin{equation}
\chi=\frac{1}{2}\sum_{\mathbf{x}_{i}}\left(\tau_{i}-\tau_{i}^{o}\right)^{2}.\label{eq:travel-time-misfit}
\end{equation}
It should be noted that $\tau_{i}$ depends on the model flow through
a complicated, as yet unknown relation $\tau_{i}\left(\mathbf{v}\left(\mathbf{x}\right)\right)$,
and similarly $\tau_{i}^{o}$ depends on the reference supergranule
velocity $\mathbf{v}^{\mathrm{ref}}\left(\mathbf{x}\right)$. Requiring
the misfit $\chi$ to decrease at each step leads to a scheme for
updating $\mathbf{v}\left(\mathbf{x}\right)$. A fundamental question
still lingers --- is reducing the travel-time misfit equivalent to
converging to the correct supergranular flow? To investigate this,
we define model misfit $\chi_{m}$ as the normalized $L_{2}$ norm
of the difference between the true model $m^{\mathrm{ref}}$ and iterated
model $m$, 
\begin{equation}
\chi_{m}=\frac{\int_{B}d\mathbf{x}\left|m^{\mathrm{ref}}\left(\mathbf{x}\right)-m\left(\mathbf{x}\right)\right|^{2}}{\int_{B}d\mathbf{x}\left|m^{\mathrm{ref}}\left(\mathbf{x}\right)\right|^{2}},\label{eq:mod-mis}
\end{equation}
where $B$ denotes the computational box, and the model $m$ represents
the variable we are interested in, for example the vector potential
$\psi$, or the individual components of the velocity $\mathbf{v}$.
Another related diagnostic measure is the model misfit as a function
of depth, defined as 
\begin{equation}
\chi_{m,z}\left(z\right)=\frac{\int_{-L/2}^{L/2}dx\left|m^{\mathrm{ref}}\left(\mathbf{x}\right)-m\left(\mathbf{x}\right)\right|^{2}}{\int_{-L/2}^{L/2}dx\left|m^{\mathrm{ref}}\left(\mathbf{x}\right)\right|^{2}},\label{eq:mod_mis_dep}
\end{equation}
where $L$ is the horizontal extent of the computational box. This
profile of model misfit allows us to probe the accuracy of the reconstructed
flow profile as a function of depth, something that can be analyzed
for each radial mode.

\section{Numerics}

\begin{figure*}
\begin{centering}
\includegraphics[scale=0.5]{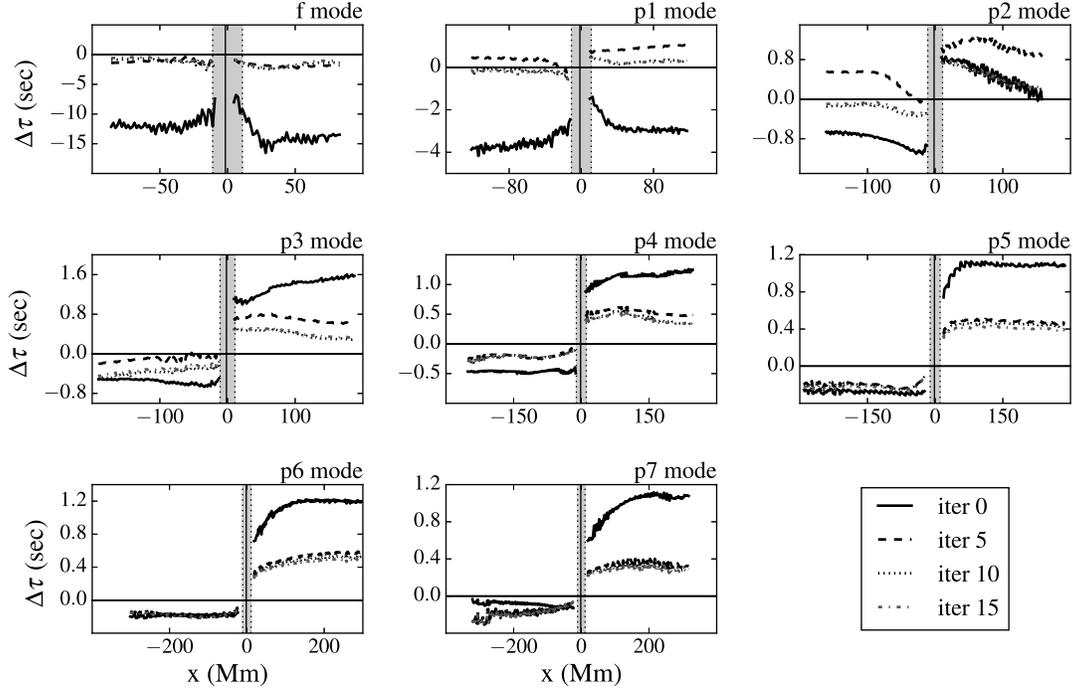}
\par\end{centering}

\caption{Travel-time difference between true and iterated models, measured
at each receiver for one particular source located at $x=-2$ Mm (solid
vertical line). The vertical patch centered about $x=0$ Mm represents
the location of the reference supergranule. Waves being sped-up would
record a negative travel-time shift whereas waves that are being slowed
down would produce a positive shift. The first few iterations are
dominated by the $f$ and $p_{1}$ modes, while the higher $p$ modes
contribute later on, although to a limited extent.\label{fig:Travel-times-receiver}}
\end{figure*}
\begin{figure*}
\begin{centering}
\includegraphics[scale=0.55]{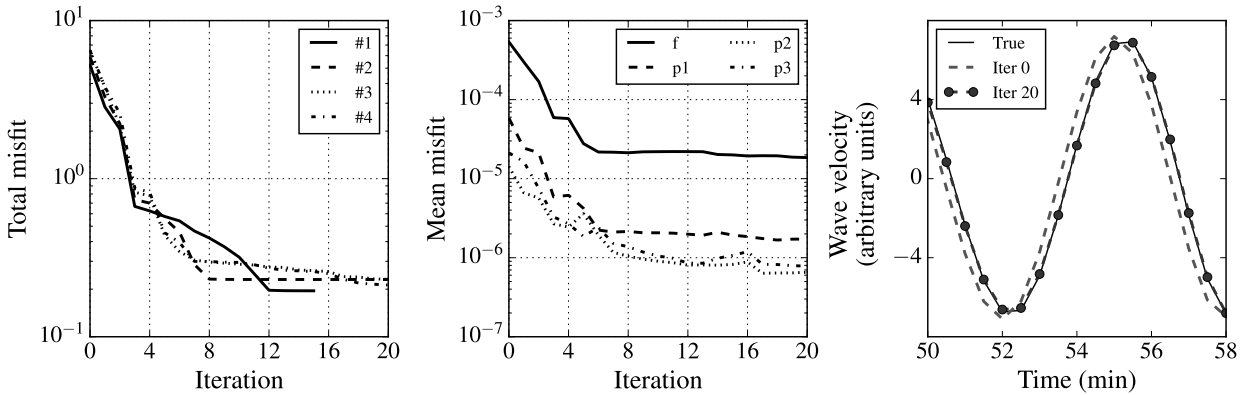}
\par\end{centering}

\caption{Left: data misfit $\chi$ (Equation (\ref{eq:travel-time-misfit}))
--- the variable that is minimized at each step --- for each strategy
(See Table \ref{tab:Strategies}). We find that data misfit falls
fastest and the most in Strategy $\#1$, where we use $f$ and $p_{1}$
modes. Middle: average modal travel-time misfit per source-receiver
pair for strategy $\#3$, where we use modes $f$ through $p_{7}$.
Only the four lowest radial orders are plotted. Right: A section of
$f$-mode waveforms at the start and after $35$ iterations for strategy
$\#3$, for a source at $x=-2$ Mm and a receiver at $x=18$ Mm. \label{fig:data_misfit}}

\end{figure*}
\begin{figure*}
\begin{centering}
\includegraphics[scale=0.45]{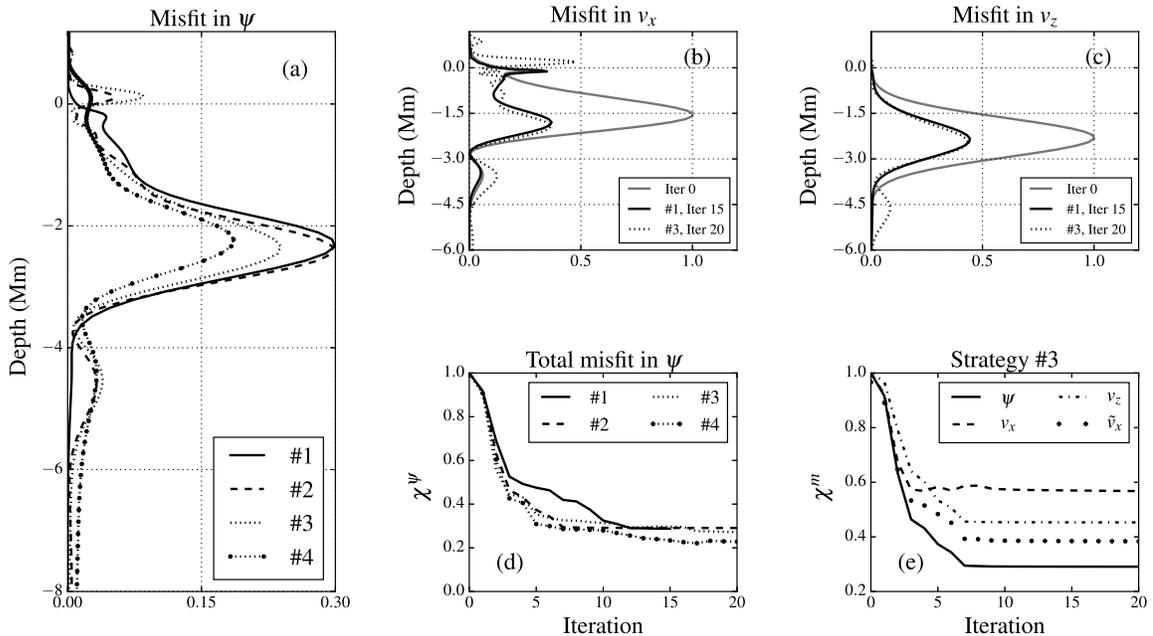}
\par\end{centering}

\caption{Model misfits in the various strategies. (a) Horizontally integrated
misfit in vector potential $\psi$ as a function of depth (Equation
(\ref{eq:mod_mis_dep}) with $m=\psi$) for strategies $\#1$ through
$\#4$; (b) horizontally integrated misfit in the horizontal component
of velocity $v_{x}$ as a function of depth; (c) horizontally integrated
misfit in the vertical component of velocity $v_{z}$ as a function
of depth; (d) total model misfit for $\psi$ (Equation (\ref{eq:mod-mis})
with $m=\psi$) as a function of iterations for strategies $\#1$
through $\#4$; (e) model misfit for vector potential (solid line),
vertical velocity component $v_{z}$ (dot-dashed line), horizontal
velocity component $v_{x}$ (dashed line), $v_{x}$ considering only
the sub-surface layers (dots, denoted by $\tilde{v}_{x}$ in the legend)
with iteration for Strategy $\#3$, that uses modes $f$ through $p_{7}$.
\label{fig:model-misfit}}
\end{figure*}
\begin{figure*}
\begin{centering}
\includegraphics[scale=0.44]{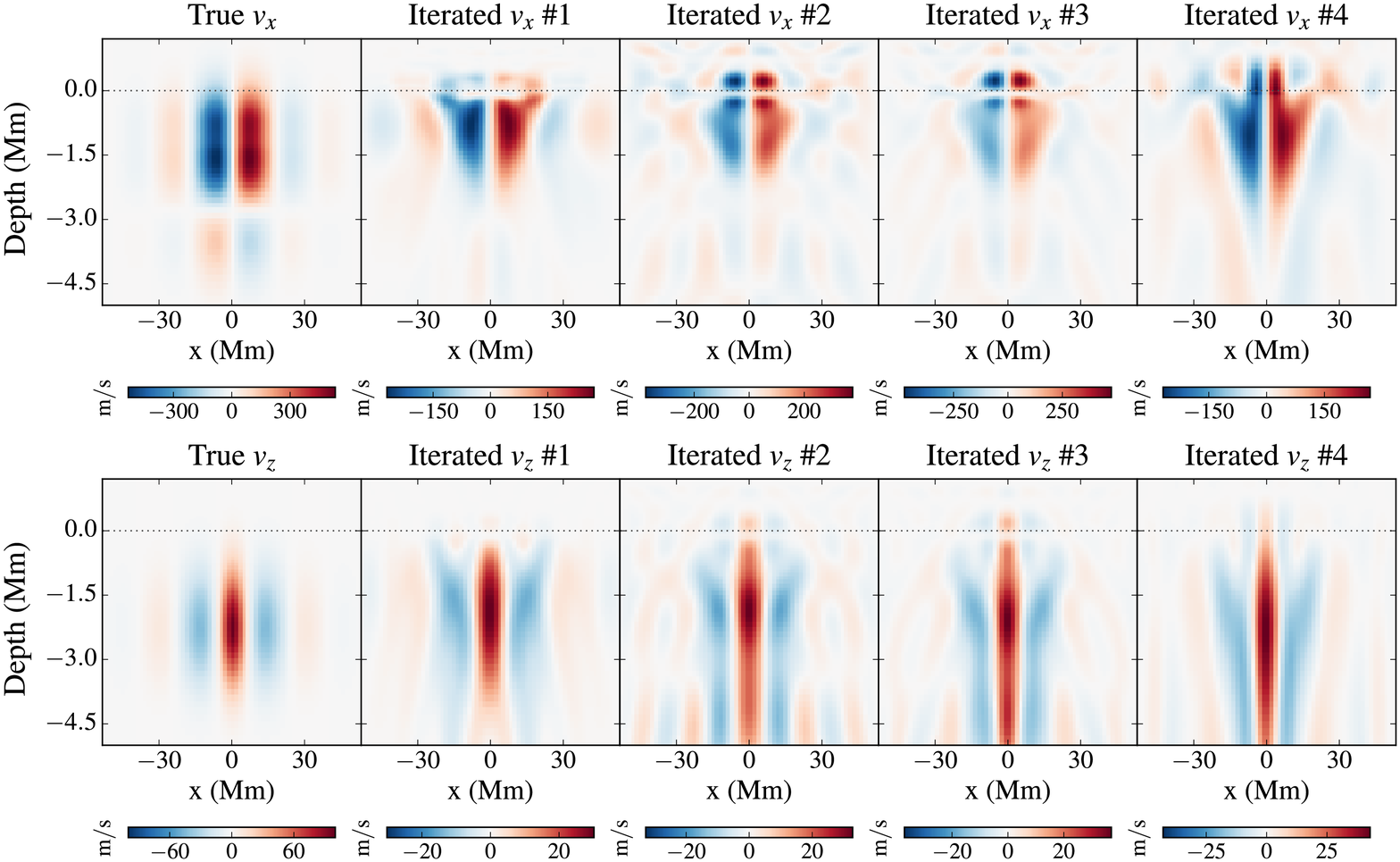}
\par\end{centering}

\caption{The leftmost panels represent the horizontal component of flow (top)
and the vertical component (bottom) of the reference supergranule.
Panels second from left onwards show the iterated velocity profile
corresponding to strategies $\#1$ through $\#4$. The horizontal
dotted line indicates the surface ($z=0$ Mm). \label{fig:True-and-iterated-flow}}
\end{figure*}
We use the wave propagation code SPARC \citep{han07phd} to carry
out simulations and measure source-receiver travel-times. SPARC uses
a second-order, five-stage low dissipation and dispersion Runge-Kutta
time-evolution scheme \citep{hu96}, where spatial derivatives are
computed using spectral (horizontal) and sixth-order compact finite-difference
\citep[vertical;][]{lele92} schemes. We choose a grid that spans
$800$ Mm horizontally resolved using $512$ points, and vertically
extends from $137$ Mm below the photosphere to $1.18$ Mm above it
over $300$ points. The computational grid is assumed to be periodic
in the horizontal direction. The vertical grid points are equally
spaced in acoustic radius, with perfectly-matched absorbing layers
lining the upper and lower boundaries of the grid to minimize reflections
back into the box. 

Lacking a specific choice for the sub-surface profile of the supergranule,
we start from a model that does not have any flows and iteratively
update it. Flow velocity is related to the vector potential through
\begin{equation}
\mathbf{v}=\frac{1}{\rho}\grad\times\left[\rho c\left(\psi-\psi_{0}\right)\mathbf{e}_{y}\right],
\end{equation}
where $\psi_{0}$ is a constant and represents the value of $\psi$
far away from the cell center, therefore choosing $\psi=\psi_{0}$
at the first step ensures that we do not have flows in our model.
At each step of the iteration, we perform eight simulations with sources
placed at different horizontal locations, all at a depth of $150$
km below the photosphere. We choose the horizontal coordinate of the
sources lying around the center of the box so that the waves traverse
a large distance through the supergranule, therefore contributing
significantly to the travel-time misfit. We carry out measurements
of the wave velocity at various horizontal locations at a height of
$200$ km above the photosphere; the horizontal range of receiver
pixels is limited by the distance traveled by the wave in the time
duration of the simulation. One particular arrangement of a source
and receiver position with respect to the reference supergranular
flow is depicted in Figure \ref{fig:flow_src}.

Waves emanating from a source reaches different receivers at different
times, and keeping track of the arrival at each receiver, we can construct
a time-distance diagram that encodes all information about wave travel-times
as measured on the surface. An example of such a time-distance diagram
is given in Figure \ref{fig:Time-distance-diagram}, where separate
arrivals show up as distinct ridges. We carry out spatial and temporal
Fourier transforms of the time-distance diagram to obtain the spectrum
of seismic oscillations measured at the surface. One such spectrum
is shown in Figure \ref{fig:spectrum}, where several ridges corresponding
to standing modes have been marked. Waves corresponding to each mode
travel at a specific phase velocity; we apply ridge-filters to isolate
them and measure travel times for each ridge. We choose several different
combinations of spectral\textbf{ }ridges as listed in Table \ref{tab:Strategies}
to carry out the inversion and compare the eventual inferences. We
apply the adjoint method --- described by \citet{han11} in the context
of helioseismology --- to construct the requisite sensitivity kernel
to update our models of the vector potential. We describe these kernels
in the appendix. The kernel represents the gradient of the misfit
functional $\chi$ with respect to the vector potential $\psi$; we
precondition the kernel using an ``approximate Hessian'' \citep{ficht11,zhu13}.
We smooth the kernels horizontally and vertically to remove high-spatial-frequency
variations, this acts as a regularization by moderating the absolute
values of high order derivatives. We apply a Gaussian smoothing: we
use a spectral smoothing horizontally with a standard deviation of
$4$ wavenumbers, and a spatial smoothing vertically with widths as
listed in Table \ref{tab:Strategies}. These kernels are used to update
our model using the Polak--Ribi\`ere variant of the non-linear conjugate
gradient scheme \citep{Nocedal2006NO}. 

\begin{table}
\begin{centering}
\begin{tabular}{|c|c|c|}
\hline 
\multirow{1}{*}{Strategy} & \multirow{1}{*}{Modes} & \multicolumn{1}{c|}{Vertical Smoothing}\tabularnewline
\hline 
$\#1$ & $f$ and $p_{1}$ & $2$ pixel\tabularnewline
\hline 
$\#2$ & $f$, $p_{1}$ to $p_{3}$ & $2$ pixel\tabularnewline
\hline 
$\#3$ & $f$, $p_{1}$ to $p_{7}$ & $2$ pixel\tabularnewline
\hline 
$\#4$ & $f$, $p_{1}$ to $p_{7}$ & $8$ pixel\tabularnewline
\hline 
\end{tabular}
\par\end{centering}

\caption{Strategies used to generate and regularize sensitivity kernels used
in the inversion.\label{tab:Strategies}}
\end{table}

Waves corresponding to different modes contribute differently to the
travel-time misfit. We show the variation of travel-time misfit as
measured at each receiver in Figure \ref{fig:Travel-times-receiver}.
We compare the reduction in data misfit for the different strategies
in Figure \ref{fig:data_misfit}. We find that the data misfit corresponding
to each radial order falls by about an order of magnitude. The corresponding
waveforms are well matched and further iterations do not improve the
misfit significantly. A question that we seek to address is --- to
what extent does a decrease in data misfit translate to an improvement
in the supergranule flow profile? We compare the improvement in model
misfit for each strategy in Figure \ref{fig:model-misfit}. We find
that strategy $1$ is able to capture many essential structural features
of the flow profile, including the return flow, whereas incorporating
higher order $p$-modes modes leads to updates concentrated at the
surface and above it. We plot the iterated velocity profiles corresponding
to the different strategies in Figure \ref{fig:True-and-iterated-flow}.
We observe that the iterations try to reach a compromise between the
flow profile above and below the surface. As a result, including higher
order $p$-modes does not lead to a significant improvement in the
profile. The misfit reduces somewhat in strategy $\#4$ --- where
we increase the degree of smoothing --- since this limits the weight
of the kernel towards deeper layers; however, better strategies that
constrain updates to sub-surface layers need to be explored. We reach
the following conclusions: (a) the deep return flow present in the
reference supergranule is not correctly reconstructed in the iterations,
indicating limited sensitivity at greater depths, (b) the peak magnitude
of the flow components are underestimated by the inversion process,
(c) the depth corresponding to maximum flow velocity is underestimated
(d) the flow velocity the surface is overestimated. There is improvement
beyond the results by \citet{dom13} in the sense that the vertical
component of the flow resembles the reference model, at least qualitatively,
however the magnitudes of the velocities are mismatched.

A similar analysis can be carried out for sound-speed perturbations,
and has been presented in the Appendix. This serves as a contrast
to the inversion for supergranular flow, and lets us compare the rates
of convergence in the two cases for the same strategies. We find that
the inversion for sound-speed perturbation converges faster and more
accurately than that of the supergranular flow, but estimates a shallower
and weaker perturbation similar to that for the flow. This seems to
hint at a shortcoming of the inversion process.

\section{Discussion}

Estimating small-scale sub-surface flows in the Sun is a fundamental
challenge of local helioseismology. Supergranular flows have proven
hard to image accurately, and despite considerable efforts, a consensus
regarding the optimal strategy has remained elusive. In this work,
we tested the effectiveness of full-waveform inversion in the context
of sub-surface supergranular flows. We generated synthetic data by
simulating seismic waves in presence of an axisymmetric supergranule.
We used travel-times of seismic waves measured on the surface to define
a misfit functional that encodes the difference between the reference
supergranule and a test model of the flow profile. We applied the
adjoint method \citep{han11} to construct kernels and iteratively
improved our model using the nonlinear conjugate-gradient algorithm
to reduce the travel-time misfit. We compared several inversion strategies
by varying the number of eigenmodes involved in the inversion as well
as the level of regularization. We found that
\begin{itemize}
\item including high order $p$-modes focuses the inversion on surface layers
at the expense of sub-surface layers. A similar near-surface deviation
was found even for a sound-speed perturbation. The inferences might
improve by imposing a surface constraint on the quantity being inverted
for.
\item the reconstructed vector potential has smoother gradients than the
true model and hence the magnitudes of velocity are underestimated.
\end{itemize}
One possible factor that is keeping the iterations from converging
onto the true model is the high dimension of the model space. In this
work we have sought kernels defined on a grid of $1.5\times10^{5}$
points. Smoothing reduces the number of independent parameters, but
the inversion is not well constrained --- the updates are not localized
to the spatial extent of the reference supergranule. A better approach
might be to expand the supergranule in an appropriate basis, an approach
similar to global seismic inversions for solar rotation. This might
lead to better convergence, and will be explored in future work. One
extra simplification that we carried out in this analysis was that
we used two-dimensional simulations to construct kernels, leaving
out scatterings off the $x-z$ plane. All such scatterings carry different
information about the flow profile --- even if the reference model
that we are trying to invert for is axisymmetric --- and therefore
might improve the convergence. 

In this work we have not split waves into frequency regimes to study
their sensitivities to the flow. A frequency filtered approach to
inversion is common in terrestrial seismology, where lower frequencies
are introduced first and higher ones gradually during the iterative
inversion, with the order being important. Lower frequencies for a
specific radial order probe deeper layers than higher frequencies,
therefore such an approach might be useful infer these layers first;
however, the range of frequencies in helioseismology is limited to
a a factor of $3$, whereas the frequency range spans orders of magnitude
in geophysics. Notwithstanding this, an approach that separates out
spatial and frequency scales is worth pursuing to obtain better inferences. 

Our analysis assumes that there is no noise in the measured wave velocity,
and therefore the travel-times measurements are precise, but this
is far from what is expected in reality. \citet{han11} prescribed
a strategy to construct a misfit functional that includes a noise
covariance matrix; a similar full-waveform inversion analysis with
such a definition of misfit would represent what we might expect in
solar observations. 

JB would like to acknowledge the financial support provided by Department
of Atomic Energy, India. 

\appendix

\section{Kernels and Updates\label{appendix:Computing-Kernels}}

We study the source-receiver travel-time for waves belonging to different
modes, ranging from $f$ through $p_{7}$. Waves corresponding to
each mode sense different regions in the solar interior, and hence
contributes differently to the inversion. The response of waves to
a small change in the model of sub-surface flows is represented by
$\mathbf{K_{v}\left(\mathbf{x}\right)}=\left(K_{v_{x}}\left(\mathbf{x}\right),K_{v_{z}}\left(\mathbf{x}\right)\right)$,
where the components describe the sensitivity of seismic waves to
horizontal and vertical components of flow velocity respectively.
The functional representation of $\mathbf{K_{v}\left(\mathbf{x}\right)}$
in terms of the wave-field can be obtained by using the adjoint method;
the procedure has been described by \citet{han11}. The kernel for
flow velocity can be manipulated to obtain the analogous kernel corresponding
to the vector potential $\psi$ , denoted by $K_{\psi}$ \citep{han14}.
The two kernels are related through
\begin{equation}
K_{\psi}=\rho c\psi\mathbf{e}_{y}\cdot\grad\times\frac{\mathbf{K_{v}}}{\rho}.
\end{equation}

\begin{figure*}
\begin{centering}
\includegraphics[scale=0.5]{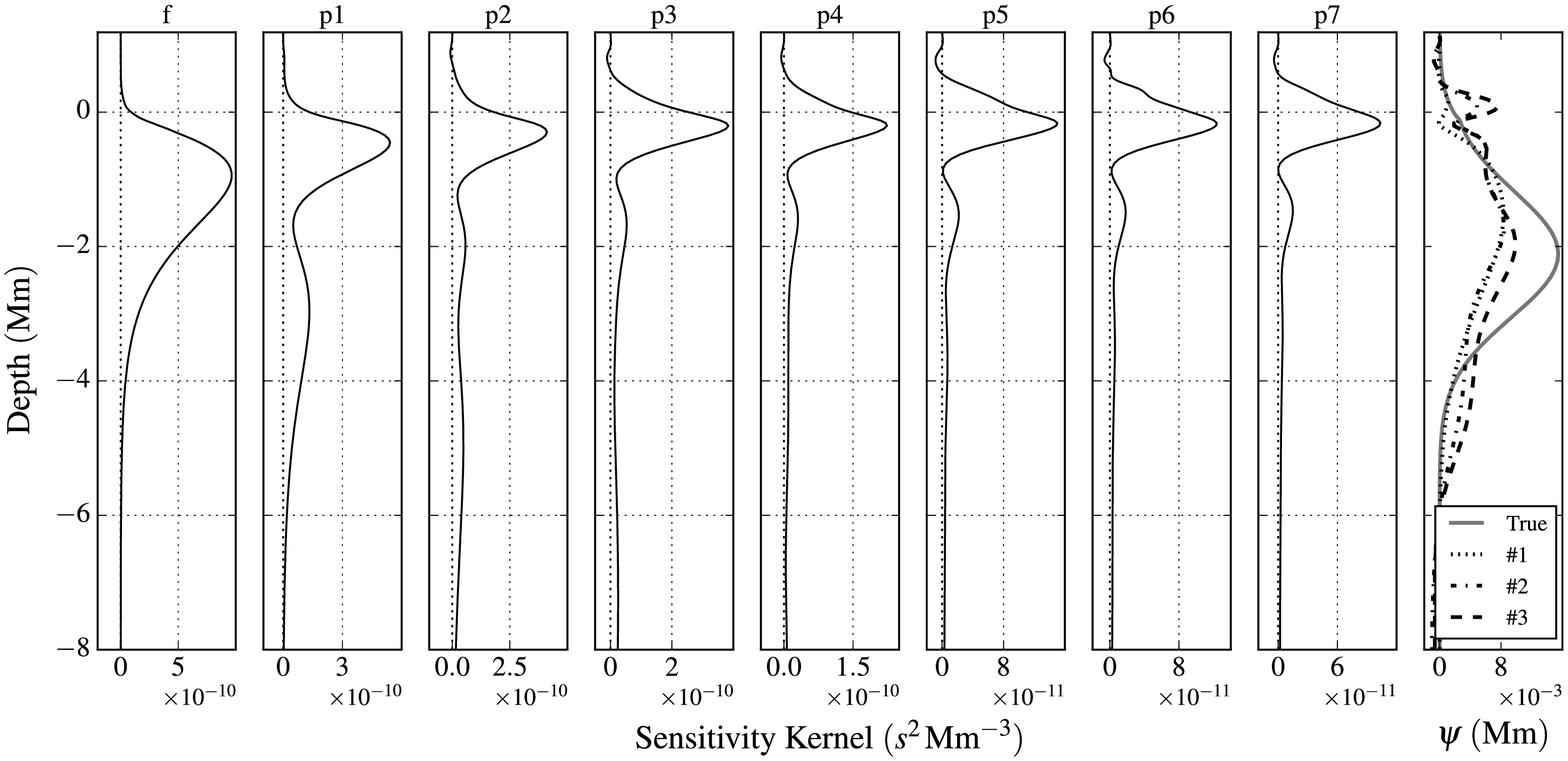}
\par\end{centering}

\caption{Horizontally integrated sensitivity kernels corresponding to modes
of different radial orders, for the vector potential $\psi$ (eight
panels from the left). Rightmost panel: True and inverted vector potentials
contrasted. \label{fig:kernel}}

\end{figure*}

We have plotted vector-potential kernels corresponding to several
modes in Figure \ref{fig:kernel}. The high $p$-mode kernels have
sharp peaks at the surface, but are sensitive to deeper layers inside
the Sun. The key to an inversion using high $p$-modes therefore seems
to be to fix surface layers first --- possible by starting with $f$
and $p_{1}$ modes --- so that the depth-sensitivity of these modes
is not masked by near-surface corrections. The strategy seems contradictory
to the geophysics technique of introducing low frequencies first,
so a conclusion has to be deferred.

We validated the flow kernels for single source-receiver travel-times
by comparing two models: one being that of a quiet sun, the other
model included a spatially constant flow $\delta v_{x}=1\,\mathrm{m/s}$.
For a single source at point $\mathbf{x}_{s}$ and receiver at point
$\mathbf{x}_{r}$, the travel-time misfit can be expressed as 
\begin{equation}
\chi=\frac{1}{2}\left(\tau_{r}-\tau_{r}^{o}\right)^{2},
\end{equation}
and a small change in model would shift this misfit by 
\begin{equation}
\delta\chi=\left(\tau_{r}-\tau_{r}^{o}\right)\delta\tau_{r}.
\end{equation}
We can express this in terms of the sensitivity kernel for the flow
as 
\begin{eqnarray}
\delta\chi & = & \left(\tau_{r}-\tau_{r}^{o}\right)\frac{\partial\tau_{r}}{\partial\mathbf{v}}\delta\mathbf{v}\nonumber \\
 & = & \int d\mathbf{x}\,\mathbf{K_{v}}\left(\mathbf{x}\right)\cdot\delta\mathbf{v}.\label{eq:kernel_verification}
\end{eqnarray}
We can measure $\delta\chi$ by comparing the waveforms arriving at
the receiver point $\mathbf{x}_{i}$ in the two models. We verified
that the equality in Equation (\ref{eq:kernel_verification}) was
maintained for a range of distances, thereby validating the kernels.

Once we have the kernels, we update our model as follows \citep{Nocedal2006NO,ficht11}:

Given $x_{0}$, evaluate $f_{0}=f\left(x_{0}\right),\;\grad f_{0}=\grad f\left(x_{0}\right)$.
Set $p_{0}\leftarrow-\grad f_{0},\;k\leftarrow0$. 

while $\grad f_{k}\neq0$

\qquad{}Evaluate $\alpha_{k}$ using a line-search

\qquad{}Evaluate $x_{k+1}=x_{k}+\alpha p_{k}$; $\grad f_{k+1}=\grad f\left(x_{k+1}\right)$

\qquad{}Precondition using ``approximate Hessian'' $\grad f_{k+1}\leftarrow H\grad f_{k+1}$

\qquad{}Compute the following: 
\begin{eqnarray*}
\beta_{k+1} & \leftarrow & \frac{\grad f_{k+1}^{T}\left(\grad f_{k+1}^{T}-\grad f_{k}^{T}\right)}{\left\Vert \grad f_{k}\right\Vert ^{2}},\\
p_{k+1} & \leftarrow & -\grad f_{k+1}+\beta_{k+1}p_{k},\\
k & \leftarrow & k+1,
\end{eqnarray*}

end while.

\section{Sound-speed perturbation\label{appendix:Sound-speed-perturbation}}

\begin{figure*}
\begin{centering}
\includegraphics[scale=0.55]{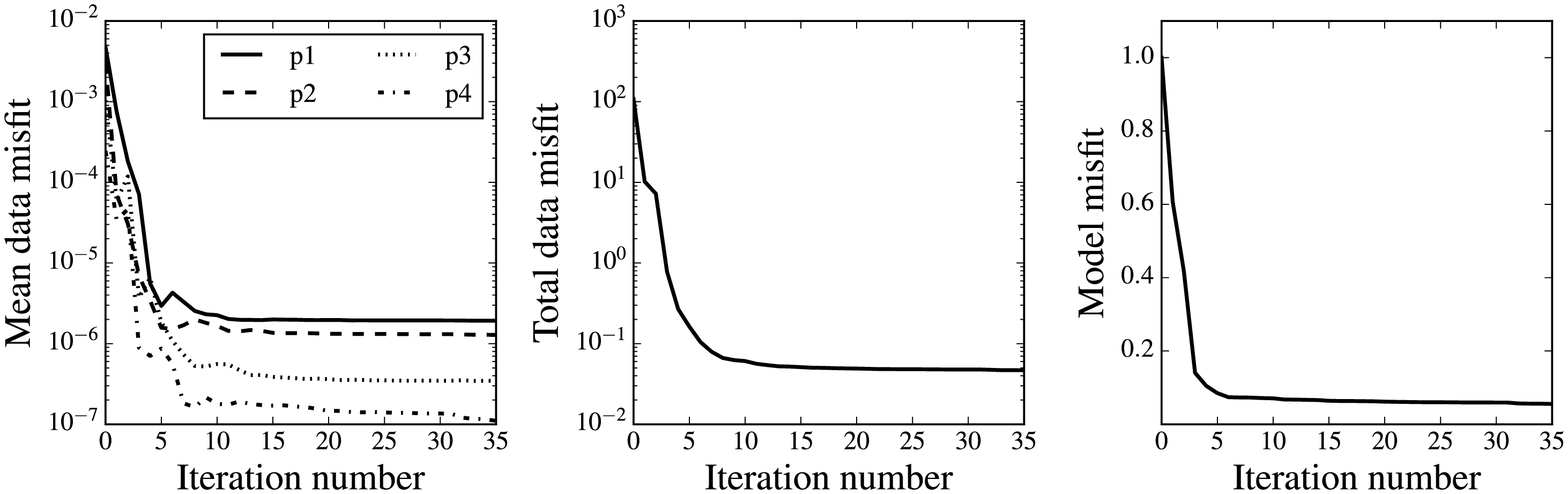}
\par\end{centering}

\caption{Data misfit for the sound-speed perturbation described in Equation
\ref{eq:ss_pert}. Left: Mean data misfit per source-receiver pair
for four of the seven radial mode used in the inversion. Center: Total
data misfit summed over all sources, receivers and modes ($\chi$
as in equation (\ref{eq:travel-time-misfit})). Right: model misfit
($\chi_{m}$ as in Equation (\ref{eq:mod-mis}), with $m=\delta\ln c$).
We find that the inversion converges by iteration $10$, further iterations
leading to tiny improvements to the model.\label{fig:misfit_ss}}

\end{figure*}

\begin{figure*}
\begin{centering}
\includegraphics[scale=0.5]{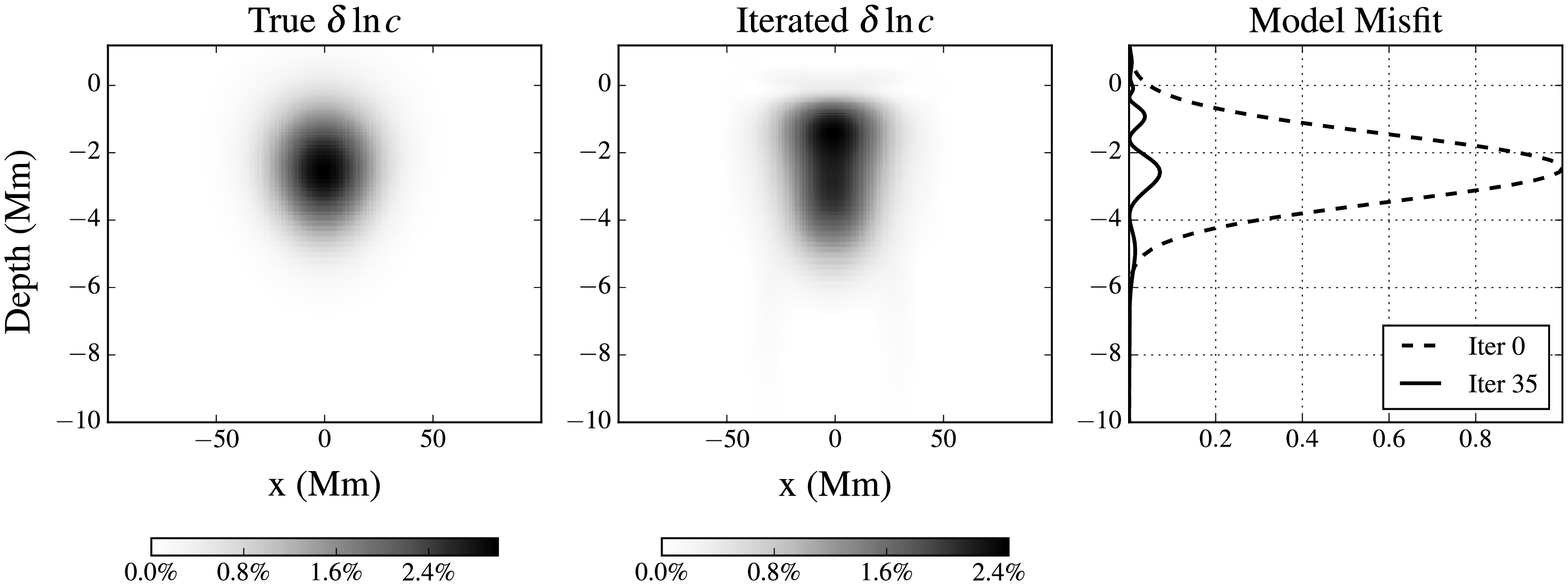}
\par\end{centering}

\caption{Left: True sound-speed perturbation. Center: Inverted sound-speed
perturbation after $35$ iterations. Right: horizontally integrated
model misfit ($\chi_{m,z}$ as in Equation (\ref{eq:mod_mis_dep}),
with $m=\delta\ln c$).\label{fig:true-iter-ss}}
\end{figure*}

We carry out an analysis similar to the supergranule flow profile
by starting with a specific perturbation to the sound-speed profile,
with a functional form given by 
\begin{equation}
\delta\ln c=\exp\left(-\frac{x^{2}}{2\sigma_{x}^{2}\left(z\right)}-\frac{\left(z-z_{0}\right)^{2}}{2\sigma_{z}^{2}}\right),\label{eq:ss_pert}
\end{equation}
where $z_{0}=-2.5$ Mm, $\sigma_{z}=1.4$ Mm, and 
\begin{equation}
\sigma_{x}\left(z\right)=10\,\mathrm{Mm}\left(1+\exp\left(\frac{z}{10\,\mathrm{Mm}}\right)\right).
\end{equation}
We use modes $p_{1}$ to $p_{7}$ to carry out the inversion, leaving
out the $f$ mode since it is not significantly affected by a sound-speed
perturbation. The data and model misfits after $35$ iterations have
been plotted in Figure \ref{fig:misfit_ss}. We find that the model
misfit falls to $0.06$, corresponding to a decrease in data misfit
by a factor of around $10^{3}$. True and iterated sound-speed perturbations
are plotted in Figure \ref{fig:true-iter-ss}. The important conclusions
that arise from this analysis are: (a) the data misfit for this sound-speed
perturbation falls by three orders of magnitude, as opposed to one
order for supergranular flows, (b) the inversion converged faster
for this sound-speed perturbation than that for the supergranular
flow.

\bibliographystyle{apj}
\bibliography{flow_references}

\end{document}